\documentclass[twocolumn]{aastex631}
\usepackage[utf8]{inputenc}
\usepackage{hyperref}
\usepackage[caption=false]{subfig}
\usepackage{float}
\usepackage{graphicx}
\usepackage{longtable}
\usepackage{tablefootnote}
\usepackage{threeparttable}

\begin{document}


\newcommand{\Siena}{Department of Physics and Astronomy, Siena College, 515 Loudon Rd, Loudonville, NY 12211, USA}
\newcommand{\UCB}{Department of Astronomy, University of California, Berkeley, 501 Campbell Hall 3411, Berkeley, CA, 94720, USA}
\newcommand{\seti}{SETI Institute, Mountain View, California}
\newcommand{\UWL}{Department of Physics, University of Wisconsin - La Crosse, 1725 State Street, La Crosse, WI 54601, USA}
\newcommand{\USQ}{University of Southern Queensland, Toowoomba, QLD 4350, Australia}
\newcommand{\KZA}{University of Malta, Institute of Space Sciences and Astronomy}
\newcommand{\PWJD}{The Breakthrough Initiatives, NASA Research Park, Bld. 18, Moffett Field, CA, 94035, USA}
\newcommand{\COL}{Department of Astronomy, Columbia University, 550 West 120th Street, New York, NY 10027, USA}
\newcommand{\Curtin}{International Centre for Radio Astronomy Research, Curtin University, Bentley, WA 6102, Australia}

\correspondingauthor{Noah Franz}
\email{nr25fran@siena.edu \\noahfranz13@gmail.com}

\author[0000-0003-4537-3575]{Noah Franz}
\affiliation{\UCB}
\affiliation{\Siena}

\author[0000-0003-4823-129X]{Steve Croft}
\affiliation{\UCB}
\affiliation{\seti}

\author[0000-0003-2828-7720]{Andrew P.\ V.\ Siemion}
\affiliation{\UCB}
\affiliation{\seti}
\affiliation{\KZA}

\author{Raffy Traas}
\affiliation{\UCB}
\affiliation{\UWL}

\author[0000-0002-7461-107X]{Bryan Brzycki}
\affiliation{\UCB}

\author[0000-0002-8604-106X]{Vishal Gajjar}
\affiliation{\UCB}

\author[0000-0002-0531-1073]{Howard Isaacson}
\affiliation{\UCB}
\affiliation{\USQ}

\author[0000-0002-7042-7566]{Matt Lebofsky}
\affiliation{\UCB}

\author{David H.\ E.\ MacMahon}
\affiliation{\UCB}

\author[0000-0003-2783-1608]{Danny C.\ Price}
\affiliation{\UCB}
\affiliation{\Curtin}

\author[0000-0001-7057-4999]{Sofia Z. Sheikh}
\affiliation{\UCB}

\author{Julia De Marines}
\affiliation{\UCB}

\author{Jamie Drew}
\affiliation{\PWJD}

\author{S. Pete Worden}
\affiliation{\PWJD}

\title{The Breakthrough Listen Search for Intelligent Life:\\Technosignature Search of Transiting TESS Targets of Interest}

\begin{abstract}
The Breakthrough Listen Initiative, as part of its larger mission, is performing the most thorough technosignature search of nearby stars. Additionally, Breakthrough Listen is collaborating with scientists working on NASA’s Transiting Exoplanet Survey Satellite (TESS), to examine TESS Targets of Interest (TOIs) for technosignatures. Here, we present a $1-11$\,GHz radio technosignature search of 61 TESS TOIs that were in transit during their Breakthrough Listen observation at the Robert C.\ Byrd Green Bank Telescope. We performed a narrowband Doppler drift search with a minimum S/N threshold of 10, across a drift rate range of $\pm 4$\,Hz\,s$^{-1}$, with a resolution of 3\,Hz. We removed radio frequency interference by comparing signals across cadences of target sources. After interference removal, there are no remaining events in our survey, and therefore no technosignature signals-of-interest detected in this work. This null result implies that at L, S, C, and X bands, fewer than 52\%, 20\%, 16\%, and 15\%, respectively, of TESS TOIs possess a transmitter with an equivalent isotropic radiated power greater than a few times $10^{14}\,\textrm{W}$.
\end{abstract}


\section{Introduction}
The Search for Extraterrestrial Intelligence (SETI) seeks an answer to the age-old question: Are we alone in the universe? The modern search for technosignatures, or signs of intelligent extraterrestrial life, began in the 1960s \citep{project_ozma}. Due to the limited technology available at the time, this search was restricted to 1420\,MHz, which was hypothesized to be a good candidate for a universal communication frequency. However, as technology has developed, technosignature searches have become much more advanced and can cover much wider bandwidths and larger numbers of targets. 

The Breakthrough Listen (BL) Initiative, launched in 2015, will search over 1 million targets for technosignatures over its 10-year lifetime \citep{Worden:2017}. BL operates at optical and radio wavelengths, using a wide variety of telescopes including the Robert C. Byrd Green Bank Telescope (GBT) in West Virginia, the Automated Planet Finder (APF) in California, and the CSIRO Parkes `Murriyang' 64-m radio telescope in Australia. This work presents a technosignature search of the frequency range $1 - 11$\,GHz using the GBT. The BL backend on the GBT is capable of simultaneously delivering billions of frequency channels across several GHz of bandwidth. \cite{macmahon:18} and \cite{lebofsky:19} provide information about the instrument, data formats, and post-observation data management. 

\begin{figure*}
    \centering
    \includegraphics[width=2\columnwidth]{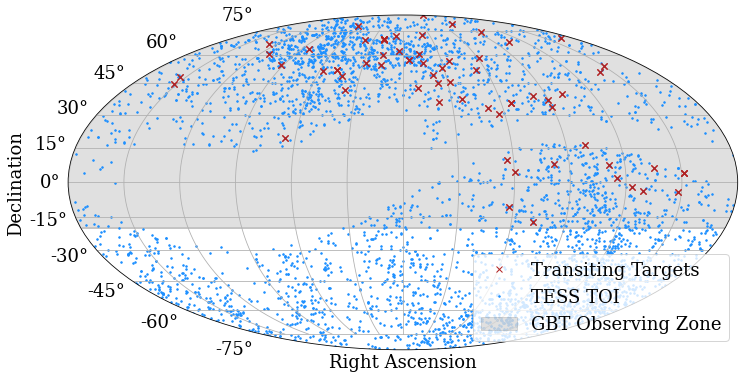}
    \caption{Sky map of all TESS TOIs (blue dots), with transiting TESS targets analyzed in this paper overlaid as red x's. The grey shaded region is the zone above declination $-20 \degr$ where BL targets are usually observed with the GBT; BL targets below this declination are usually observed at the Parkes Observatory.}
    \label{fig:sky_map}
\end{figure*}

BL employs a variety of strategies for target prioritization. One is to select targets from catalogs compiled by NASA's Transiting Exoplanet Survey Satellite (TESS). As of 2021 June, TESS has found 4,190 new exoplanets, including confirmed exoplanets and candidates, some of which may have suitable conditions for life. \cite{Traas} recently performed a technosignature search of 28 TESS targets of interest (TOIs) using the L, S, C, and X-band receivers at the GBT. Transiting systems are prioritized because Earth is in the ecliptic for these systems. An ETI may be more likely to send bright signals out in the direction of their ecliptic, either to intentionally signal observers who can see their transits, or for purposes such as interplanetary radar \citep{Traas}.

We refine the search of \citet{Traas} by selecting systems that were observed with the GBT {\em during} transits of candidate exoplanets, which may further improve the chance of receiving an extraterrestrial signal. An ETI may choose to broadcast signals towards their anti-stellar point,  knowing that observers may be monitoring their system during a transit, so there is a higher likelihood of a transmission being received. In addition, by choosing to broadcast at this special temporal ``Schelling Point'' \citep{Sheikh, 2018arXiv180906857W, Gajjar-2021}, an ETI could enhance signal detectability for a given transmitter power (relative to an omnidirectional transmitter) by increasing their antenna gain and beaming a signal in the opposite direction to their star. 

\section{Observations}

 BL targets at GBT are observed with an ``on/off" \mbox{ABACAD} cadence method \citep{lebofsky:19}. The primary target A, is observed, then an ``off" target B, is observed. This method is then repeated twice more with the same ``on" target and two new ``off" targets, C and D. Each target in the cadence is observed for 5 minutes such that the ``on" target is observed for a total of 15 minutes and each ``off'' target is observed for 5 minutes. Comparing the ``on'' and ``off'' scans allows us to differentiate between radio frequency interference (RFI) signals and a candidate ETI signal, since the latter is expected to be localized on the sky. 

\subsection{Target Selection}

\begin{figure*}[t]
    \centering
    \includegraphics[width=2\columnwidth]{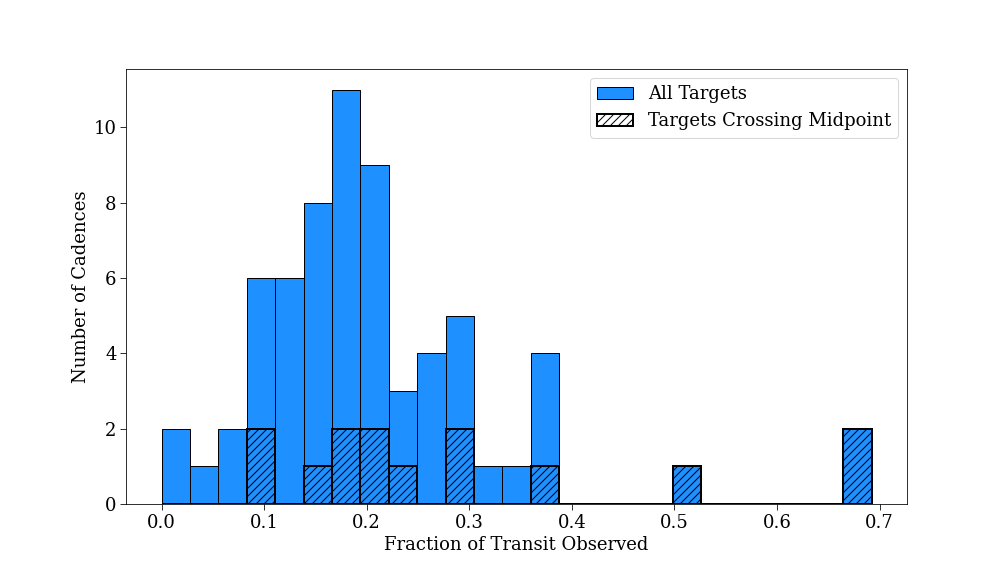}
    \caption{Fraction of the transit observed in the GBT observations. The blue bins represent all 66 cadences, while the hashed bins indicate a target that crosses the midpoint of its transit during the observation.}
    \label{fig:frac_transit}
\end{figure*}
\begin{figure*}[t]
    \centering
    \includegraphics[width=2\columnwidth]{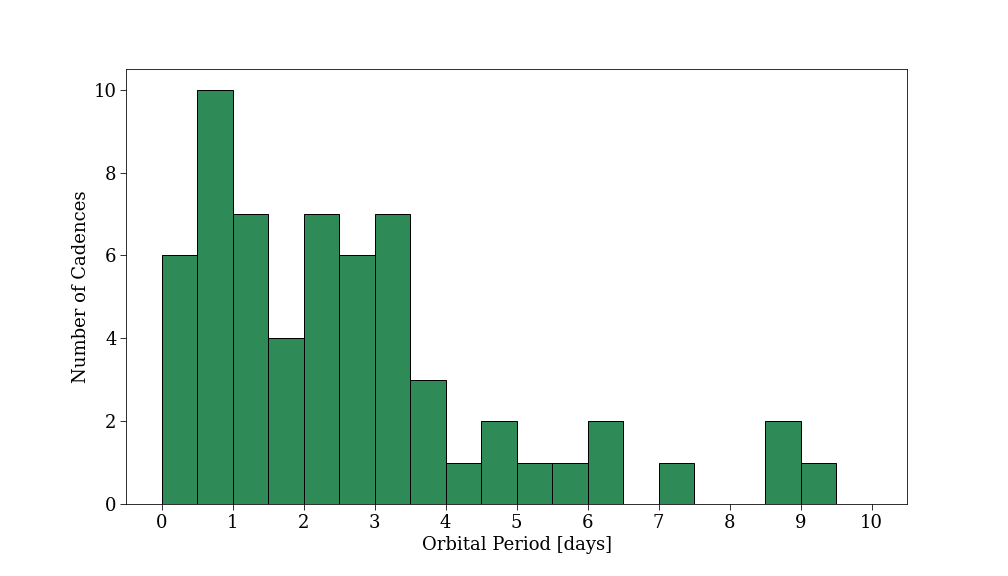}
    \caption{Histogram of the orbital periods of the observed targets.}
    \label{fig:period_hist}
\end{figure*}

Observations of TOIs by BL at the GBT are scheduled automatically by selection from target lists, and not typically deliberately timed to coincide with transits. By examining ephemerides from ExoFOP-TESS \citep{exofop} for all targets observed by BL at GBT as of 2021 June, we determined\footnote{Code at: \url{https://github.com/noahfranz13/BL-TESSsearch}} that 61 unique targets, across 66 observations, serendipitously transit during their GBT observation. These 61 targets are shown in Figure \ref{fig:sky_map} and Appendix \ref{app:targets}. TIC\,344926234 and TIC\,365683032 were observed with two different receivers during two different transits, TIC\,376637093 was observed with three different receivers during three different transits, and TIC\,286561122 was observed at C-Band twice during a single transit. The notch filter regions \citep{lebofsky:19} at L ($1200 - 1340$\,MHz) and S ($2300 - 2360$\,MHz) bands are excluded from our analysis.

A histogram of the fraction of each transit observed is shown in Figure \ref{fig:frac_transit}. The fraction of transit observed was calculated by dividing the observation time of the entire cadence by the total transit time of the exoplanet candidate,

\begin{equation}
    \textrm{FTO} = \frac{t_{\textrm{obs,~transit}}}{(t_{\textrm{egress}} - t_{\textrm{ingress}})},
    \label{eq:FTO}
\end{equation}

\noindent where FTO stands for the Fraction of Transit Observed, $t_{\textrm{obs,~transit}}$ is the amount of time in the overlap of the transit time and observation time, and $t_{\textrm{egress}}$ and $t_{\textrm{ingress}}$ are the time of egress and ingress, respectively. Targets that cross the midpoint of their transit, as shown by the hashed bins in Figure \ref{fig:frac_transit}, are especially interesting: a narrow-beamed transmitter pointing away from the host star, perhaps located at the second Lagrange point, would appear strongest at the midpoint of transit. 

Figure \ref{fig:period_hist} shows a histogram of the orbital periods of the TESS TOIs chosen for this project. These periods are all relatively short, so the TOIs are unlikely to be terrestrial planets in the habitable zone. Still, ETI may assume it is easier for us to detect these closer, short period, exoplanets and place a transmitter there.

\begin{center}
\begin{table*}
\begin{threeparttable}
    \caption{Survey Parameters}
    \begin{tabular}{c c c c c c c c}
        \hline
        \hline
        Receiver & Frequency & Cadences & Hits & Events & CWTFM\tnote{a} & $10\sigma$  EIRP$_{\textrm{min}}$  & Transmitter \\
        & [GHz] & & & & & [TW]\tnote{b} &  Limit [\%]\tnote{c} \\
        \hline
        L & 1.10 - 1.90 & 5 & 213097 & 172 & 3793 & 167 & 52\\
        S & 1.80 - 2.80 & 17 & 160057 & 33 & 2828 & 393 & 20\\
        C & 4.00 - 7.80 & 21 & 578264 & 57 & 3060 & 788 & 16\\
        X & 7.80 - 11.20 & 23 & 1503241 & 372 & 3686 & 719 & 15\\
        \hline
        Total & 1.10 - 11.20 & 66 & 2442347 & 634 & - & - & - \\
        \hline
    \end{tabular}
    \begin{tablenotes}
        \footnotesize
        \item[a] Continuous Waveform Transmitter Figure of Merit (CWTFM) is a figure of merit that describes the likelihood to find a signal above the EIRP$_{\textrm{min}}$ for that receiver.
        \item[b] Minimum Equivalent Isotropic Radiated Power (EIRP$_{\textrm{min}}$) is a measure of the minimum necessary omnidirectional power of a transmitter at each receiver to be detected.
        \item[c] The transmitter limit is the maximum percentage of exoplanet candidates in each frequency range that possess a transmitter.
    \end{tablenotes}
    \label{tab:receivers}
\end{threeparttable}
\end{table*}
\end{center}

\section{Doppler Search}

We perform our analysis on fine-frequency resolution spectrograms from the BL backend at the GBT. As described by \citet{lebofsky:19} and \citet{macmahon:18}, the BL backend records spectral data in 187.5 MHz frequency chunks, with each chunk sent to a separate compute node. Data recorded before early 2021 were spliced together in frequency, one file per receiver, before archiving. Starting in early 2021, files were instead left in their unspliced form on the compute nodes, which enables easier parallel processing. The 66 cadences analyzed here represent 21\,TB of data in total, most of which were analyzed \textit{in situ} on the GBT BL compute nodes. In one observation in our sample, TIC\,365781372 at X-band, the blc40 compute node failed to record data during a scan, leading to a gap of 187.5\,MHz in the spectrum.

Each cadence was analyzed using the BL \texttt{turboSETI} pipeline \citep{turboSETI}. First, \texttt{FindDoppler} identifies narrow-band Doppler-drifting signals in the filterbank files. Following from \citet{Price:2020} and \citet{Traas}, we adopt a minimum\footnote{\texttt{turboSETI}'s dechirping efficiency is lower for high drift rate signals, resulting in a higher effective S/N limit. For more details see \citet{Gajjar-2021}. } S/N threshold of 10, across a drift rate range of $\pm 4$\,Hz\,s$^{-1}$. To maximize efficiency, we parallelized the processing across all 64 compute nodes available to BL at GBT, greatly reducing runtime for large amounts of data.

We use the measured orbital periods for our TOIs, applying the methods presented by \citet{Sheikh-2019}, to calculate theoretical maximum drift rates for transmitters in the systems in our sample. We neglect any contribution from the rotation rates of the planets  (which are unknown, but in many cases may be negligible, since many of our targets have small periods and are most likely tidally locked). We find that only 2.4\% of our targets have {\em maximum} drift rates that lie within $\pm 4$\,Hz\,s$^{-1}$, suggesting that a search over a larger drift rate range would be optimal, albeit more computationally expensive. However, it would be simple (and maybe even common) for ETI to correct for their drift rate when transmitting a signal, so received signals would only have small drift rates due to Earth's orbit and rotation \citep{Sheikh-2019, Horowitz:1993}. Additionally, \texttt{turboSETI} will pick out bright signals even if the drift rate is not matched correctly. 

The second part of the Doppler search is to run the \texttt{find\_event} pipeline which removes signals with no drift rate and compares the hits across each cadence, eliminating any signals present in both the ``on" and ``off" observations. \texttt{find\_event} returns \textit{events}, which are any signals that are present in the ``on" and not ``off" observations. Selecting signals that are only present in the ``on" observations removes RFI and isolates signals that are localized on the sky.

Finally, the \texttt{plot\_event} pipeline produces cadence plots for visual inspection, which allows us to manually eliminate any RFI remaining after the \texttt{find\_event} pipeline. For more information see \cite{Enriquez} and \cite{turboSETI}.

\section{Results}

\begin{figure*}
    \centering
    \includegraphics[width=\textwidth]{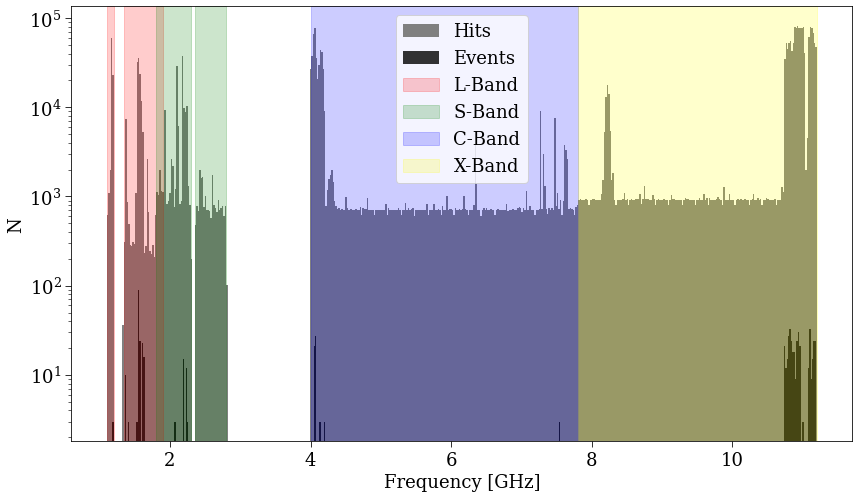}
    \caption{Number of hits (grey) and events (black) versus frequency. The frequency range (band) for each GBT receiver is represented by the colored regions; the (unshaded) notch filter regions \citep{lebofsky:19} at L and S bands are excluded from our analysis. Values for each band, with the number of cadences, are shown in Table~\ref{tab:receivers}. Note that there are a different number of cadences at each band so the number of hits and events plotted here should not be directly compared across bands.}
    \label{fig:freq_hist}
\end{figure*}

\subsection{Technosignature Search}
For the rest of this discussion, we refer to a ``hit" as any signal present in a single observation and an ``event" as a collection of related hits that successfully passed through the \texttt{find\_event} pipeline. We find 2,442,347 hits and 634 events which were distributed across the receiver bands as shown in Figure~\ref{fig:freq_hist} and Table~\ref{tab:receivers}. We show examples of events in Figure \ref{fig:int_events}. After visually inspecting all 634 events, we find that all of them are consistent with human-generated RFI. Most commonly, these signals appear to be present --- but not detectable by \texttt{turboSETI} above the S/N threshold --- throughout the entire cadence, indicating a source of interference that is likely local to the telescope. 

\newcommand{\fw}{0.95}

\begin{figure*}
\begin{center}
\qquad
\subfloat[]{
  \includegraphics[width=\fw\columnwidth]{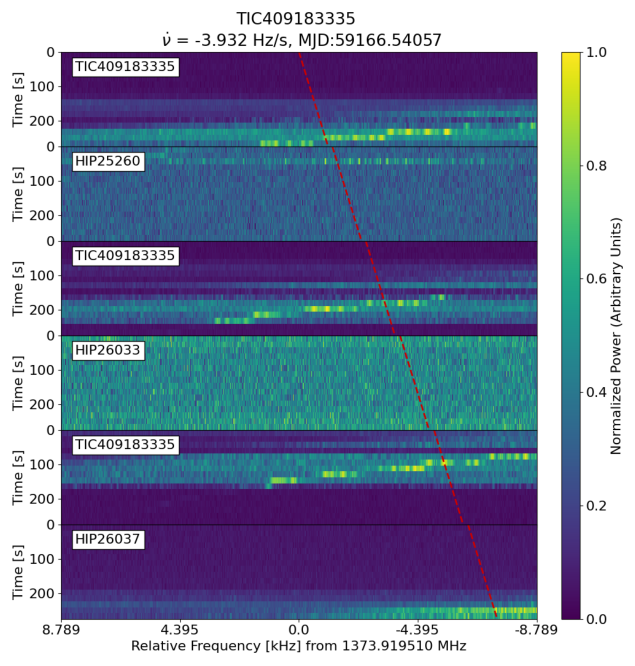}
  \label{fig:int_event2}
}
\qquad
\subfloat[]{
  \includegraphics[width=\fw\columnwidth]{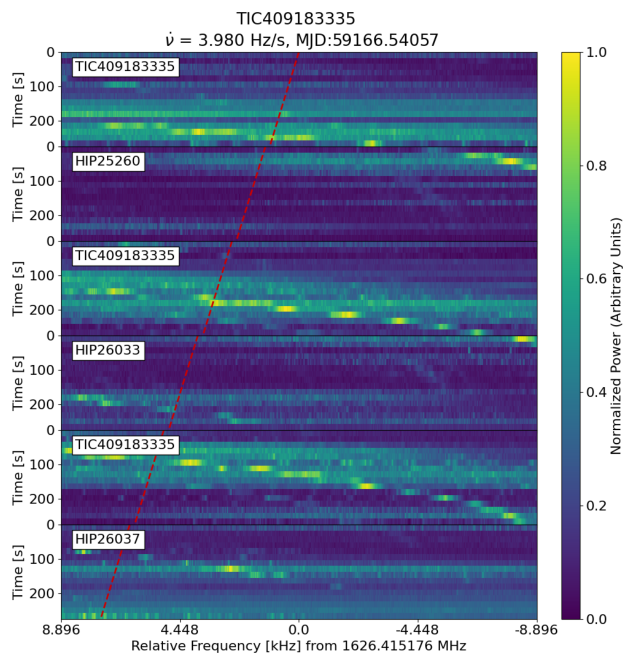}
  \label{fig:int_event3}
}

\qquad
\subfloat[]{
  \includegraphics[width=\fw\columnwidth]{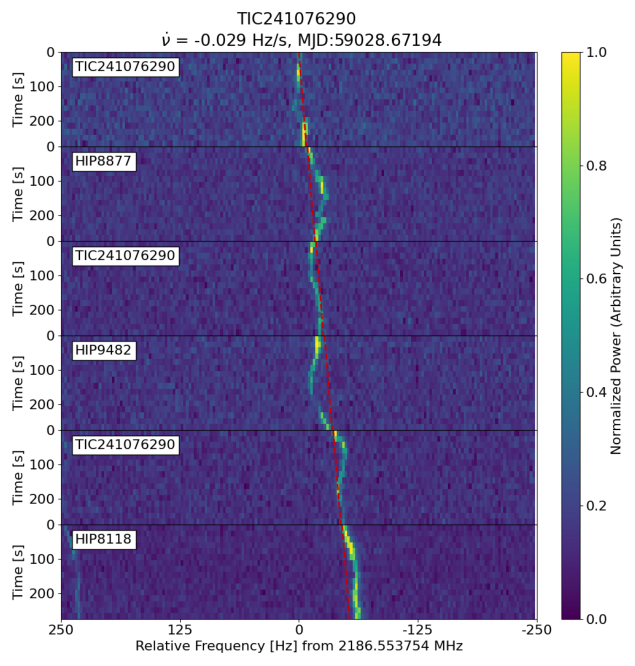}
  \label{fig:int_event1}
}
\qquad
\subfloat[]{
  \includegraphics[width=\fw\columnwidth]{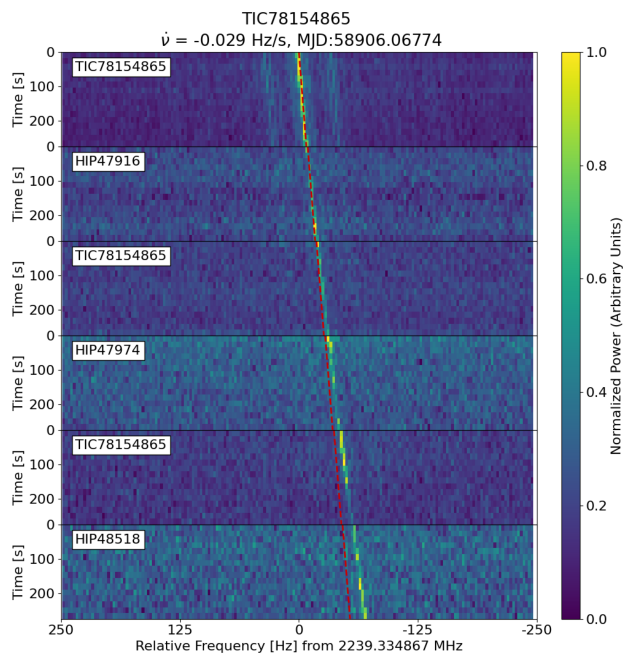}%
  \label{fig:int_event5}
}
\end{center}
\caption{Dynamic spectra (waterfall plots) of 4 representative events from the 634 event sample. Each plot is a vertical stack of the 6 scans making up an ABACAD cadence. The vertical axis shows the time since the start of each scan in the cadence, and the horizontal axis shows the frequency offset from the event's starting frequency.}
\label{fig:int_events}
\end{figure*}

Figures~\ref{fig:int_event2} and \ref{fig:int_event3} illustrate signals that seem to appear mostly in the ``on" observations in a given cadence. However, both cadences also have some similar signals in the ``off" observations, and can therefore be ruled out as signals-of-interest. These signals are broader in frequency than the narrowband drifting tones \texttt{turboSETI} is designed to search for. Nevertheless, they were bright enough to rise above \texttt{turboSETI}'s S/N threshold and register as hits. Although an ETI could transmit signals with a range of bandwidths, the broader signals in this study were clearly due to RFI. Furthermore, the signal in Figure~\ref{fig:int_event2} is in a frequency range commonly used for aeronautical radar (as are many of the top-ranked events presented by \citealt{Enriquez}). Likewise, Figure~\ref{fig:int_event3}, given its frequency, is RFI that is likely related to the Iridium satellite constellation. 

Figure~\ref{fig:int_event1} shows a waterfall plot of TIC\,241076290, a candidate exoplanet with a tight orbit around its host star, with a period of ~0.258 days. This is the only target in our analysis whose transit is shorter than the 30 minute observation. In this case we observe only the end of the transit. In the future, by scheduling specifically timed observations for systems with short transits, we could look for signals that appear only during the transit. As of 2021 June, in the ExoFOP-TESS catalog there are 31 TOIs with transits shorter than 30 minutes, which corresponds to 0.74\% of TOIs. These TOIs would be interesting targets for follow up observations. 

Figure~\ref{fig:int_event5} appears to have a non-linear Doppler shift, suggesting it is accelerating with respect to the telescope, as might be expected for a satellite in Earth orbit, and its frequency corresponds to a known satellite downlink frequency. However, due to the relative motion of satellites (even geosynchronous satellites) with respect to sidereal targets, they usually appear in only one or two scans. Instead, Figure~\ref{fig:int_event5} has a signal present throughout the entire cadence. Its presence in the ``off'' scans rules it out as an ETI candidate; it may be a pernicious example of a slow-moving satellite (possibly visible through a telescope sidelobe) that was moving in the same general direction as the telescope over the course of the 30-minute observation.

\subsection{Hit and Event Distribution}

\begin{figure*}
    \centering
    \includegraphics[width=2\columnwidth, height=0.88\textheight]{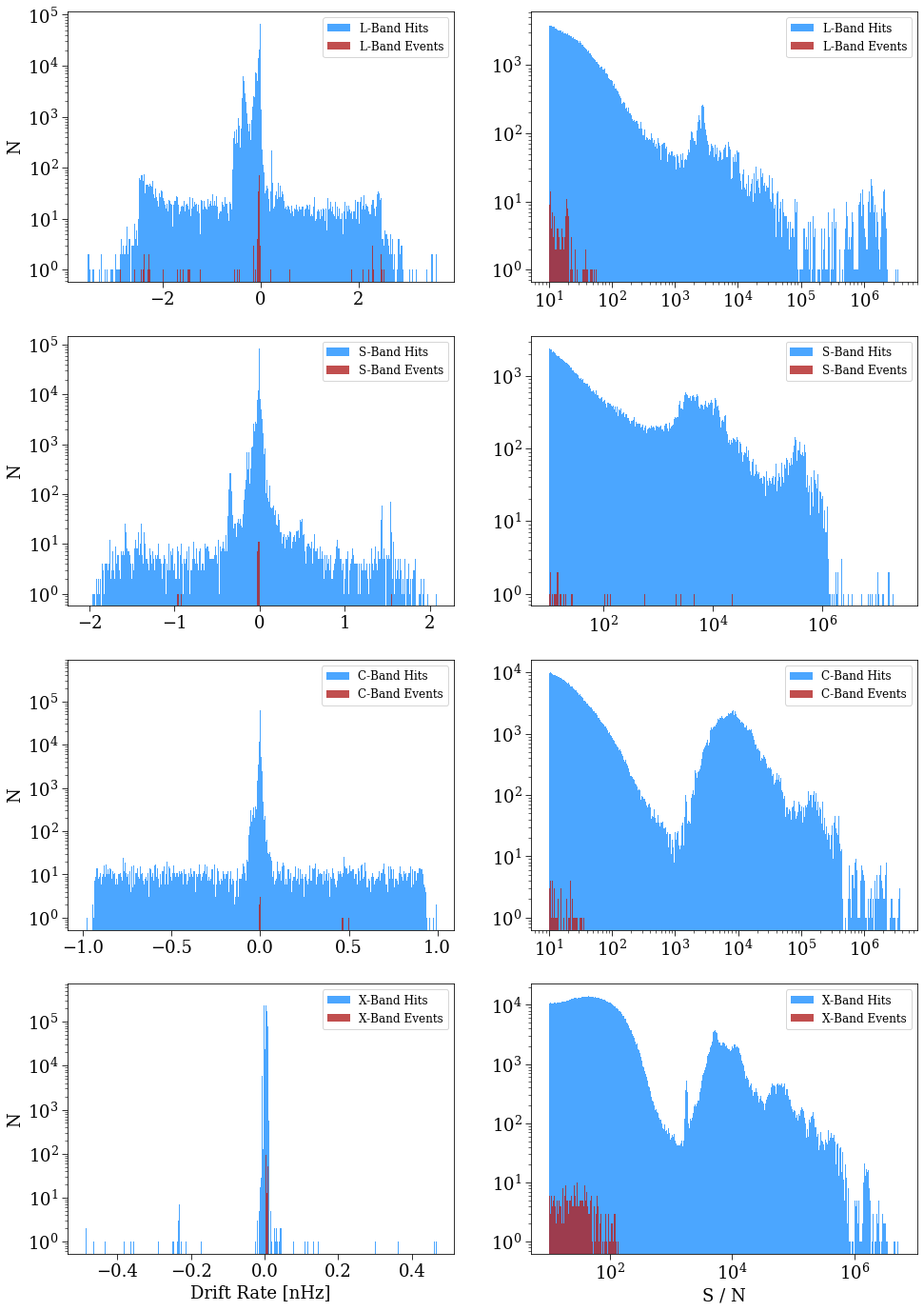}
    \caption{{\em Left Column:} Histograms of hits (blue) and events (red) as a function of drift rate split up by receiver. The reported drift rate is normalized by the center frequency of the hit or event, to produce a value in units of nHz (e.g., 1\,nHz = 1\,Hz\,s$^{-1}$ at 1\,GHz). {\em Right Column:} Histograms of hits (blue) and events (red) as a function of S/N split up by Green Bank Telescope receiver. Note that not all receivers observed the same number of targets.}
    \label{fig:hit_dist}
\end{figure*}

The hit and event frequency distributions are shown in Figure~\ref{fig:freq_hist}. Histograms of the S/N and drift rate distributions are shown in Figure~\ref{fig:hit_dist}. There are significantly more hits and events at low drift rates, likely produced by RFI local to the telescope.  

\subsection{Figures of Merit}

To further evaluate our ability to detect ETI signals in this work, we can compare our figures-of-merit to those from past SETI studies. One such figure-of-merit is the Drake Figure of Merit  \citep[DFM;][]{Drake:1984},

\begin{equation}\label{eq:DFM}
    \textrm{DFM} = \frac{n~\Delta f~\Omega}{F^{3/2}_{\textrm{min}}},
\end{equation}

\noindent where $n$ is the number of observations at a receiver, $\Delta f$ is the total frequency range observed, $\Omega$ is the full width half maximum of the receiver, and $F_\textrm{min}$ is the minimum detectable flux. While DFM has some limitations, as discussed by \citet{Enriquez} and \citet{Margot-2021}, it is still a useful statistic, especially for surveys across multiple receivers, such as this one, because it incorporates both the bandwidth surveyed and the minimum detectable power. Table~\ref{tab:DFM} shows the DFM for this study in comparison to other recent searches; larger DFMs indicate more comprehensive searches. 

\begin{table}
    \centering
    \caption{Drake Figure of Merit}
    \begin{tabular}{c c c}
        \hline
        \hline
        Study & DFM [GHz m$^3$ W$^{3/2}$] \\
        \hline
        This Study & $2.1\times10^{32}$\\
        \cite{Margot-2021} & $1.11\times10^{32}$\\
        \cite{Gajjar-2021} & $4\times10^{28}$\\
        \hline
    \end{tabular}
    \label{tab:DFM}
\end{table}

\begin{figure*}
    \centering
    \includegraphics[width=2\columnwidth]{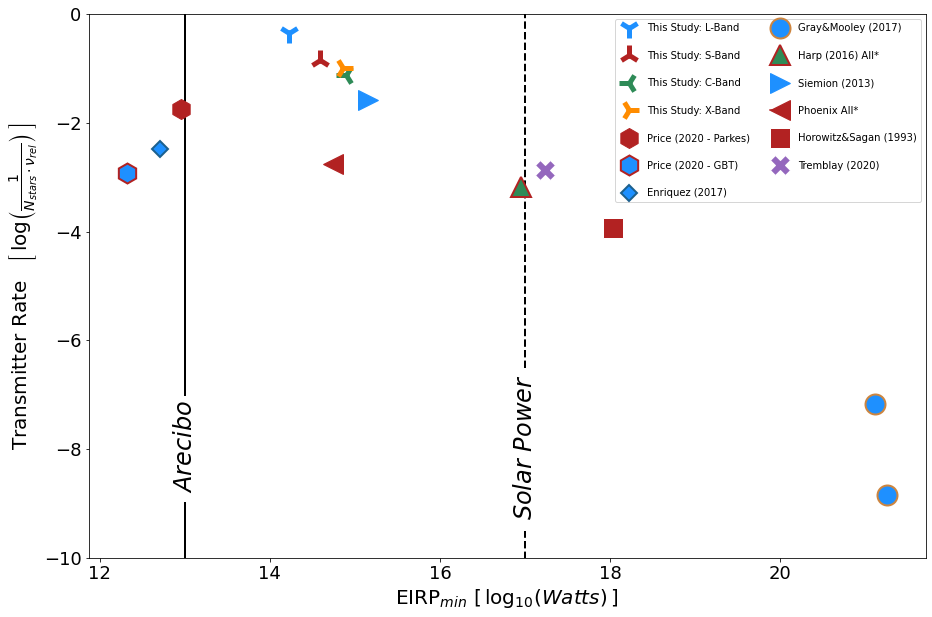}
    \caption{Transmitter Rate versus EIRP$_{\textrm{min}}$ for this study (represented by the four Y-shaped points) compared to past studies (none of which specifically targeted systems during transits). The two vertical lines represent the EIRP of the Arecibo S-Band radar, and the solar power incident on Earth.}
    \label{fig:eirp}
\end{figure*}

A second useful figure-of-merit is the Continuous Waveform Transmitter Figure of Merit \citep[CWTFM;][]{Enriquez}. This describes the likelihood of finding an ETI signal above a specific minimum Equivalent Isotropic Radiated Power (EIRP$_{\textrm{min}}$),

\begin{equation}\label{eq:CWTFM}
    \textrm{CWTFM} = \zeta_{\textrm{AO}} ~ \frac{\textrm{EIRP}_{\textrm{min}}}{N_\textrm{stars}~\nu_{\textrm{rel}}}
\end{equation}

\noindent where $N_\textrm{stars}$ is the number of pointings in a survey at a receiver times the number of stars per pointing (assumed to be 1), $\nu_{\textrm{rel}}$ is the total bandwidth for a receiver normalized by the central frequency of the receiver, and $\zeta_{\textrm{AO}}$ is a normalization constant such that CWTFM is 1 for an EIRP equal to that of Arecibo. EIRP$_{\textrm{min}}$ is a measure of the necessary power of a hypothetical omnidirectional antenna, in the most distant star system in our sample, to be detected by each GBT receiver. We plot the Transmitter Rate (CWTFM divided by EIRP$_{\textrm{min}}$) vs.\ EIRP$_{\textrm{min}}$ for our study in comparison to past searches in Figure~\ref{fig:eirp}. Technosignature searches represent compromises between sensitivity (higher sensitivity towards the left-hand side of the figure) and sky and bandwidth coverage (more stars, and/or wider fractional bandwidth coverage, towards the bottom of the figure). Our study occupies a similar region of parameter space to previous studies, but is the first to achieve wide frequency coverage for a significant number of stars observed \textit{during transit} of candidate exoplanets. 

\subsection{Transmitter Limit}

Given our lack of detection of any signal-of-interest, we can calculate the transmitter limit, or maximum percentage of TESS TOIs at each band that possess a detectable transmitter based on our search parameters. \cite{Price:2020}, \cite{Traas}, and other authors, calculate this limit using a one sided 95\% Poisson confidence interval with a 50\% probability of actually observing a signal if the transmitter is present \citep{1986ApJ...303..336G}. Given the small number of cadences observed at L-band, a binomial confidence interval is a better estimate for the transmitter limit in our case.  We list the relevant limits (95\% one-sided binomial interval, with a 50\% probability of detecting a signal if present) in Table~\ref{tab:receivers}. Work is ongoing to determine more accurate detection thresholds by performing signal injection and recovery in BL data.

The TOIs observed in this work all have short periods, as shown in Figure~\ref{fig:period_hist}. These targets are very close to their host stars, receiving many hundreds of times more stellar insolation than terrestrial planets in the habitable zone. Additionally, some of our targets are exoplanet candidates rather than confirmed exoplanets. Some caution is therefore warranted in extrapolating the transmitter limits from the TESS TOIs observed in this work to the entire population of exoplanets.

\section{Conclusions}

We performed a technosignature search of 61 TESS TOIs, over 66 observations, that are in transit during their BL observation at the GBT. This could be a favored time to search for technosignatures from ETI because Earth is in the ecliptic of these exoplanet candidates as they transit. ETIs may determine such transits are good Schelling Points, and time their transmissions accordingly. 

After searching the 66 cadences for technosignatures, we did not find any potential technosignature signals-of-interest. Using this null result, we constrain the existence of extraterrestrial transmitters brighter than a few hundred TW to less than 52\%, 20\%, 16\%, and 15\% (for L, S, C, and X bands, respectively) of TESS TOIs that are observed during transit. 

\section{Future Studies}

There are numerous ways to extend the work presented in the previous sections. First, we could analyze targets that serendipitously enter or exit their \textit{secondary} transit during their observation. This way, we could search for signals that appear or disappear as the exoplanet candidates passes behind its host star.  

Second, we could search for signals that appear or disappear at the same time as targets enter or exit their transit during a BL observation. This would require observations that cover the entire transit, including a substantial portion of data taken outside of the ingress and egress of the transit. Since ETI may assume that Earth is observing the exoplanet during its transit, they may transmit a beacon to Earth only during the transit. 

Third, we could search for signals that appear close to the midpoint of the transit as viewed from Earth. In this scenario, ETI may direct a narrow, beamed signal towards their anti-stellar point, which might appear as a signal that changes in intensity with a Gaussian shape, as Earth is swept by the transmitter beam. Such a signal could come from a transmitter present on the anti-stellar point of a tidally-locked planet or, as previously mentioned, a transmitter placed at an exoplanet's second Lagrange point.

Fourth, rather than rely on serendipitous scheduling of BL observations of TESS TOIs, observations could be scheduled during exoplanet candidate transits. This could enable larger, more thorough studies of exoplanet candidate transits as a Schelling Point, and as a geometrically-favourable region for technosignature searches.

\section{Acknowledgements}

The Breakthrough Prize Foundation funds the Breakthrough Initiatives which manages Breakthrough Listen. The Green Bank Observatory facility is supported by the National Science Foundation, and is operated by Associated Universities, Inc. under a cooperative agreement. We thank the staff at Green Bank Observatory for their support with operations. NF was funded as a participant in the Berkeley SETI Research Center Research Experience for Undergraduates Site, supported by the National Science Foundation under Grant No.~1950897. This research has made use of the Exoplanet Follow-up Observation Program website, which is operated by the California Institute of Technology, under contract with the National Aeronautics and Space Administration under the Exoplanet Exploration Program.
\newpage
NF thanks all of the 2021 Berkeley SETI interns for their support and encouragement. In addition, we thank Richard Elkins for his support with running \texttt{turboSETI} and Daniel Est\'evez for his insight in identifying the signal in Figure \ref{fig:int_event5}. We thank the anonymous referee for their comments on the manuscript.

\nocite{Phoenix_all:2004, Tremblay:2020, Horowitz:1993, Harp:2016, Siemion:2013, GrayMooley:2017dr}


\bibliographystyle{aasjournal}
\bibliography{main}

\begin{thebibliography}{}
\expandafter\ifx\csname natexlab\endcsname\relax\def\natexlab#1{#1}\fi
\providecommand{\url}[1]{\href{#1}{#1}}
\providecommand{\dodoi}[1]{doi:~\href{http://doi.org/#1}{\nolinkurl{#1}}}
\providecommand{\doeprint}[1]{\href{http://ascl.net/#1}{\nolinkurl{http://ascl.net/#1}}}
\providecommand{\doarXiv}[1]{\href{https://arxiv.org/abs/#1}{\nolinkurl{https://arxiv.org/abs/#1}}}

\bibitem[{{Backus} \& {Project Phoenix Team}(2004)}]{Phoenix_all:2004}
{Backus}, P.~R., \& {Project Phoenix Team}. 2004, in American Astronomical
  Society Meeting Abstracts, Vol. 204, American Astronomical Society Meeting
  Abstracts \#204, 75.04

\bibitem[{Drake(1961)}]{project_ozma}
Drake, F. 1961, Physics Today, 14, 40,
  \dodoi{https://doi.org/10.1063/1.3057500}

\bibitem[{{Drake} {et~al.}(1984){Drake}, {Wolfe}, \& {Seeger}}]{Drake:1984}
{Drake}, F., {Wolfe}, J.~H., \& {Seeger}, C.~L. 1984, SETI science working
  group report, Tech. rep.

\bibitem[{{Enriquez} \& {Price}(2019)}]{turboSETI}
{Enriquez}, E., \& {Price}, D. 2019, {turboSETI: Python-based SETI search
  algorithm}.
\newblock \doeprint{1906.006}

\bibitem[{{Enriquez} {et~al.}(2017){Enriquez}, {Siemion}, {Foster}, {Gajjar},
  {Hellbourg}, {Hickish}, {Isaacson}, {Price}, {Croft}, {DeBoer}, {Lebofsky},
  {MacMahon}, \& {Werthimer}}]{Enriquez}
{Enriquez}, J.~E., {Siemion}, A., {Foster}, G., {et~al.} 2017, \apj, 849, 104,
  \dodoi{10.3847/1538-4357/aa8d1b}

\bibitem[{{ExoFOP}(2019)}]{exofop}
{ExoFOP}. 2019, Exoplanet Follow-up Observing Program - TESS,  IPAC,
  \dodoi{10.26134/EXOFOP3}

\bibitem[{{Gajjar} {et~al.}(2021){Gajjar}, {Perez}, {Siemion}, {Foster},
  {Brzycki}, {Chatterjee}, {Chen}, {Cordes}, {Croft}, {Czech}, {DeBoer},
  {DeMarines}, {Drew}, {Gowanlock}, {Isaacson}, {Lacki}, {Lebofsky},
  {MacMahon}, {Morrison}, {Ng}, {de Pater}, {Price}, {Sheikh}, {Suresh},
  {Webb}, \& {Pete Worden}}]{Gajjar-2021}
{Gajjar}, V., {Perez}, K.~I., {Siemion}, A. P.~V., {et~al.} 2021, \aj, 162, 33,
  \dodoi{10.3847/1538-3881/abfd36}

\bibitem[{{Gehrels}(1986)}]{1986ApJ...303..336G}
{Gehrels}, N. 1986, \apj, 303, 336, \dodoi{10.1086/164079}

\bibitem[{Gray \& Mooley(2017)}]{GrayMooley:2017dr}
Gray, R.~H., \& Mooley, K. 2017, The Astronomical Journal, 153, 110,
  \dodoi{10.3847/1538-3881/153/3/110}

\bibitem[{{Harp} {et~al.}(2016){Harp}, {Richards}, {Tarter}, {Dreher},
  {Jordan}, {Shostak}, {Smolek}, {Kilsdonk}, {Wilcox}, {Wimberly}, {Ross},
  {Barott}, {Ackermann}, \& {Blair}}]{Harp:2016}
{Harp}, G.~R., {Richards}, J., {Tarter}, J.~C., {et~al.} 2016, \aj, 152, 181,
  \dodoi{10.3847/0004-6256/152/6/181}

\bibitem[{{Horowitz} \& {Sagan}(1993)}]{Horowitz:1993}
{Horowitz}, P., \& {Sagan}, C. 1993, \apj, 415, 218, \dodoi{10.1086/173157}

\bibitem[{{Lebofsky} {et~al.}(2019){Lebofsky}, {Croft}, {Siemion}, {Price},
  {Enriquez}, {Isaacson}, {MacMahon}, {Anderson}, {Brzycki}, {Cobb}, {Czech},
  {DeBoer}, {DeMarines}, {Drew}, {Foster}, {Gajjar}, {Gizani}, {Hellbourg},
  {Korpela}, {Lacki}, {Sheikh}, {Werthimer}, {Worden}, {Yu}, \&
  {Zhang}}]{lebofsky:19}
{Lebofsky}, M., {Croft}, S., {Siemion}, A. P.~V., {et~al.} 2019, \pasp, 131,
  124505, \dodoi{10.1088/1538-3873/ab3e82}

\bibitem[{{MacMahon} {et~al.}(2018){MacMahon}, {Price}, {Lebofsky}, {Siemion},
  {Croft}, {DeBoer}, {Enriquez}, {Gajjar}, {Hellbourg}, {Isaacson},
  {Werthimer}, {Abdurashidova}, {Bloss}, {Brandt}, {Creager}, {Ford}, {Lynch},
  {Maddalena}, {McCullough}, {Ray}, {Whitehead}, \& {Woody}}]{macmahon:18}
{MacMahon}, D. H.~E., {Price}, D.~C., {Lebofsky}, M., {et~al.} 2018, \pasp,
  130, 044502, \dodoi{10.1088/1538-3873/aa80d2}

\bibitem[{Margot {et~al.}(2021)Margot, Pinchuk, Geil, Alexander, Arora, Biswas,
  Cebreros, Desai, Duclos, Dunne, Fu, Goel, Gonzales, Gonzalez, Jain, Lam,
  Lewis, Lewis, Li, MacDougall, Makarem, Manan, Molina, Nagib, Neville,
  O'Toole, Rockwell, Rokushima, Romanek, Schmidgall, Seth, Shah, Shimane,
  Singhal, Tokadjian, Villafana, Wang, Yun, Zhu, \& Lynch}]{Margot-2021}
Margot, J.-L., Pinchuk, P., Geil, R., {et~al.} 2021, The Astronomical Journal,
  161, 55, \dodoi{10.3847/1538-3881/abcc77}

\bibitem[{Price {et~al.}(2020)Price, Enriquez, Brzycki, Croft, Czech, DeBoer,
  DeMarines, Foster, Gajjar, Gizani, \& et~al.}]{Price:2020}
Price, D.~C., Enriquez, J.~E., Brzycki, B., {et~al.} 2020, The Astronomical
  Journal, 159, 86, \dodoi{10.3847/1538-3881/ab65f1}

\bibitem[{{Sheikh} {et~al.}(2020){Sheikh}, {Siemion}, {Enriquez}, {Price},
  {Isaacson}, {Lebofsky}, {Gajjar}, \& {Kalas}}]{Sheikh}
{Sheikh}, S.~Z., {Siemion}, A., {Enriquez}, J.~E., {et~al.} 2020, \aj, 160, 29,
  \dodoi{10.3847/1538-3881/ab9361}

\bibitem[{{Sheikh} {et~al.}(2019){Sheikh}, {Wright}, {Siemion}, \&
  {Enriquez}}]{Sheikh-2019}
{Sheikh}, S.~Z., {Wright}, J.~T., {Siemion}, A., \& {Enriquez}, J.~E. 2019,
  \apj, 884, 14, \dodoi{10.3847/1538-4357/ab3fa8}

\bibitem[{Siemion {et~al.}(2013)Siemion, Demorest, Korpela, Maddalena,
  Werthimer, Cobb, Howard, Langston, Lebofsky, Marcy, \& Tarter}]{Siemion:2013}
Siemion, A. P.~V., Demorest, P., Korpela, E., {et~al.} 2013, The Astrophysical
  Journal, 767, 94

\bibitem[{{Traas} {et~al.}(2021){Traas}, {Croft}, {Gajjar}, {Isaacson},
  {Lebofsky}, {MacMahon}, {Perez}, {Price}, {Sheikh}, {Siemion}, {Smith},
  {Drew}, \& {Worden}}]{Traas}
{Traas}, R., {Croft}, S., {Gajjar}, V., {et~al.} 2021, \aj, 161, 286,
  \dodoi{10.3847/1538-3881/abf649}

\bibitem[{{Tremblay} \& {Tingay}(2020)}]{Tremblay:2020}
{Tremblay}, C.~D., \& {Tingay}, S.~J. 2020, \pasa, 37, e035,
  \dodoi{10.1017/pasa.2020.27}

\bibitem[{Worden {et~al.}(2017)Worden, Drew, Siemion, Werthimer, DeBoer, Croft,
  MacMahon, Lebofsky, Isaacson, Hickish, Price, Gajjar, \&
  Wright}]{Worden:2017}
Worden, S.~P., Drew, J., Siemion, A., {et~al.} 2017, Acta Astronautica, 139,
  98, \dodoi{10.1016/j.actaastro.2017.06.008}

\bibitem[{{Wright} {et~al.}(2018){Wright}, {Sheikh}, {Alm{\'a}r}, {Denning},
  {Dick}, \& {Tarter}}]{2018arXiv180906857W}
{Wright}, J.~T., {Sheikh}, S., {Alm{\'a}r}, I., {et~al.} 2018, NASA
  Technosignatures Workshop, arXiv e-prints, arXiv:1809.06857.
\newblock \doarXiv{1809.06857}

\end{thebibliography}

\appendix
\section{Target List}\label{app:targets}
\begin{longtable}{cccccccc}
    \hline
    \hline
Target Name  & TOI     & RA (hrs)    & Dec ($^\circ$) & Distance (pc) & Orbital  & Observation & Band \\
&&&&&Period (days)&Start Time (UTC)& \\

\hline
TIC438490744 & 529.01  & 6.683726312 & 16.58970997    & 63.0507       & 1.665878              & 5/31/2021 21:36  & C    \\
TIC147950620 & 1194.01 & 11.18809073 & 69.96478246    & 149.667       & 2.310602              & 6/14/2020 20:37  & C    \\
TIC458478250 & 1165.01 & 15.47643744 & 66.35871037    & 126.261       & 2.255296              & 11/24/2019 23:54 & S    \\
TIC344926234 & 634.01  & 10.10383058 & 3.946493286    & 92.8368       & 0.49359               & 11/24/2019 14:43 & S, C   \\
TIC78154865  & 638.01  & 9.857820581 & -4.123319468   & 96.6355       & 0.493826              & 2/27/2020 1:37   & S    \\
TIC365781372 & 627.01  & 5.461566558 & 7.918525328    & 629.834       & 1.13889               & 11/30/2019 5:03  & X    \\
TIC270468559 & 571.01  & 9.022952699 & 6.097088972    & 405.238       & 4.641843              & 9/11/2020 19:08  & C    \\
TIC121338379 & 498.01  & 8.606046796 & -3.860202132   & 188.364       & 0.275043              & 1/19/2020 8:34   & S    \\
TIC375542276 & 1163.01 & 19.60611228 & 19.63922729    & 148.342       & 3.07765               & 12/15/2019 1:12  & L    \\
TIC468880077 & 438.01  & 3.766972039 & 9.9903089      & 72.4646       & 5.8076                & 1/27/2021 23:58  & C    \\
TIC459942762 & 430.01  & 4.018554669 & 4.540889327    & 66.5727       & 0.58644               & 12/21/2019 6:56  & X    \\
TIC280437559 & 969.01  & 7.675771778 & 2.098612197    & 77.2554       & 1.823737              & 1/11/2020 9:20   & X    \\
TIC425206121 & 508.01  & 7.433963292 & 7.615772707    & 300.276       & 4.611733              & 1/19/2020 3:44   & S    \\
TIC178367144 & 966.01  & 8.226143814 & -1.982782058   & 253.985       & 3.409244              & 1/19/2020 8:02   & S    \\
TIC138168780 & 1651.01 & 6.319529079 & 73.82755828    & 235.479       & 3.764988              & 3/29/2021 18:57  & L    \\
TIC73104318  & 1674.01 & 4.114769716 & 58.46544652    & 201.645       & 7.45494               & 2/10/2020 4:40   & X    \\
TIC422756130 & 1695.01 & 1.461444537 & 72.29660211    & 45.1309       & 3.134319              & 5/25/2020 18:35  & S    \\
TIC285674856 & 1570.01 & 3.546414139 & 51.88450172    & 294.123       & 1.74626               & 2/17/2020 20:06  & X    \\
TIC241076290 & 1560.01 & 1.935519308 & 52.58547107    & 560.222       & 0.25792               & 6/28/2020 16:07  & S    \\
TIC348673213 & 1639.01 & 2.387149453 & 56.57002561    & 153.986       & 0.901465              & 4/19/2020 16:04  & C    \\
TIC292321872 & 1572.01 & 2.126808592 & 45.50016306    & 505.331       & 8.66698               & 11/13/2020 9:44  & L    \\
TIC294471966 & 1446.01 & 20.1334245  & 51.36180671    & 133.863       & 6.31719               & 6/30/2020 0:34   & C    \\
TIC409183335 & 1667.01 & 5.453849363 & 38.59745582    & 225.941       & 3.32125               & 11/13/2020 12:58 & L    \\
TIC286561122\footnote{Note that this target was observed twice at C-Band and both cadences overlap transits for this TOI.} & 1658.01 & 4.391744889 & 35.49511432    & 506.864       & 0.67994               & 3/23/2020 0:16   & C    \\
TIC311035838 & 1419.01 & 13.73959477 & 48.02856107    & 134.554       & 2.899733              & 6/20/2020 5:37   & S    \\
TIC327579226 & 1532.01 & 0.315744444 & 57.20064898    & 259.563       & 8.90592               & 11/10/2020 4:26  & X    \\
TIC365683032 & 1354.01 & 20.81199235 & 51.91068918    & 245.776       & 1.42904               & 4/18/2020 8:56   & S, X   \\
TIC241040309 & 1559.01 & 1.381155015 & 48.95536157    & 685.251       & 3.46479               & 7/18/2020 13:06  & X    \\
TIC312862941 & 1638.01 & 1.021431709 & 55.69799904    & 126.283       & 0.915094              & 4/20/2020 0:22   & C    \\
TIC137881699 & 1781.01 & 10.03817034 & 53.95082988    & 935.456       & 2.972133              & 9/11/2020 23:33  & C    \\
TIC149833117 & 1717.01 & 6.975231953 & 67.67733006    & 188.086       & 4.052173              & 6/5/2020 4:12    & S    \\
TIC368536386 & 1666.01 & 5.961338651 & 36.76580927    & 428.948       & 1.69433               & 9/14/2020 7:13   & C    \\
TIC376682699 & 1511.01 & 22.69014554 & 69.07445015    & 544.233       & 1.10264               & 11/8/2020 15:36  & X    \\
TIC376637093 & 1516.01 & 22.67230188 & 69.50372602    & 247.054       & 2.05603               & 5/19/2020 22:39  & S, C, X  \\
TIC327011842 & 1576.01 & 1.564455335 & 45.01032893    & 493.702       & 0.78424               & 7/18/2020 11:58  & X    \\
TIC44631965  & 1461.01 & 1.482420079 & 35.86484113    & 359.959       & 3.568678              & 5/29/2020 19:06  & C    \\
TIC142090065 & 1715.01 & 5.271560572 & 79.73772521    & 182.907       & 2.826937              & 9/4/2020 7:39    & X    \\
TIC198212955 & 1242.01 & 16.57021523 & 60.19589615    & 110.015       & 0.381481              & 7/29/2020 8:39   & C    \\
TIC138017750 & 1608.01 & 3.386736393 & 33.07814949    & 100.635       & 2.472722              & 10/26/2020 0:59  & S    \\
TIC26433869  & 1607.01 & 3.7876164   & 30.14950686    & 329.591       & 1.03578               & 7/3/2020 18:25   & X    \\
TIC353367071 & 1663.01 & 5.995649857 & 33.50698402    & 402.261       & 2.37532               & 9/14/2020 8:18   & C    \\
TIC272625214 & 1613.01 & 23.75456899 & 62.14267079    & 304.76        & 5.24666               & 7/27/2020 16:11  & C    \\
TIC129979528 & 1599.01 & 2.447511416 & 37.55044553    & 121.944       & 1.219868              & 9/17/2020 7:05   & X    \\
TIC341815767 & 1819.01 & 17.83467483 & 54.63614716    & 160.295       & 3.09374               & 12/21/2020 19:24 & X    \\
TIC457138169 & 1770.01 & 9.424525742 & 50.9088635     & 163.438       & 1.09254               & 8/14/2020 13:08  & C    \\
TIC371673488 & 1497.01 & 22.88221213 & 59.85095835    & 405.174       & 0.8158                & 12/20/2020 1:44  & X    \\
TIC15863518  & 1713.01 & 6.701367042 & 39.84291832    & 138.371       & 0.557201              & 12/13/2020 0:47  & X    \\
TIC389182138 & 1391.01 & 22.90899711 & 54.16180798    & 115.746       & 2.72687               & 10/10/2020 6:51  & C    \\
TIC235905185 & 1829.01 & 19.39182508 & 78.75421665    & 479.529       & 6.289555              & 12/1/2020 4:15   & X    \\
TIC191284318 & 1458.01 & 0.63819763  & 42.46306636    & 226.637       & 2.77598               & 11/10/2020 7:20  & X    \\
TIC358631536 & 1343.01 & 21.17169156 & 48.4642791     & 400.034       & 3.40304               & 12/24/2020 18:22 & S    \\
TIC274942910 & 1325.01 & 21.52843349 & 41.79747049    & 52.4946       & 1.07922               & 12/24/2020 22:09 & S    \\
TIC233720539 & 1815.01 & 18.42528104 & 63.48810973    & 617.233       & 2.55532               & 1/14/2021 1:44   & X    \\
TIC38686737  & 432.01  & 3.857704881 & -10.6140933    & 746.646       & 2.24704               & 1/14/2021 5:19   & X    \\
TIC117979455 & 422.01  & 4.786847839 & -17.25336165   & 124.504       & 0.63322               & 1/17/2021 5:29   & S    \\
TIC328167090 & 1384.01 & 22.11089078 & 55.68625098    & 235.218       & 0.71255               & 4/4/2021 13:04   & L    \\
TIC154741689 & 2170.01 & 10.95424616 & 89.08691789    & 206.368       & 9.27688               & 3/17/2021 6:56   & C    \\
TIC427730490 & 2040.01 & 23.48497742 & 71.50646786    & 144.717       & 3.86085               & 3/22/2021 20:38  & X    \\
TIC321688498 & 2290.01 & 21.43996743 & 68.64052458    & 58.0924       & 0.38623               & 3/22/2021 22:53  & X    \\
TIC393911494 & 2106.01 & 13.81189142 & 44.9117615     & 121.167       & 0.633259              & 3/28/2021 11:38  & S    \\
TIC285542903 & 2060.01 & 0.884664039 & 60.61811644    & 914.062       & 2.26584               & 4/19/2021 17:49  & C  \\
\hline
    \caption[]{Our sample of TESS TOIs that transited during their observation at the GBT.}
    \label{tab:my_label}
\end{longtable}

\end{document}